\documentclass[12pt]{article}
\usepackage{epsf}
\newcommand{\eqb}{\begin{equation}}
\newcommand{\eqe}{\end{equation}}
\newcommand{\dmb}{\begin{displaymath}}
\newcommand{\dme}{\end{displaymath}}

\newcommand{\eab}{\begin{eqnarray}}
\newcommand{\eae}{\end{eqnarray}}

\newcommand{\be}{\begin{equation}}
\newcommand{\ee}{\end{equation}}

\def\lsim{\mathrel{\raise.3ex\hbox{$<$\kern-.75em\lower1ex\hbox{$\sim$}}}}
\def\gsim{\mathrel{\raise.3ex\hbox{$>$\kern-.75em\lower1ex\hbox{$\sim$}}}}
\def\Li2{{\rm Li}_2}

\newcommand{\bmG}{\mathbf G}

\newcommand{\msbar}{$\overline{\mbox{MS}}$ }

\begin{document}

\begin{titlepage}
\begin{flushright}
MPI-PhT 2004-29 \\
HD-THEP 04-6 \\
March 2004
\end{flushright}
\vspace{0.6cm}

\begin{center}
\Large{{\bf $\overline{\rm MS}$ Charm Mass from
Charmonium Sum Rules with Contour Improvement}}

\vspace{1cm}

 A.~H.~Hoang$^a$ and M.~Jamin$^b$  

\end{center}
\vspace{0.3cm}

\begin{center}
{\em $^a$Max-Planck-Institut f\"ur Physik\\ 
Werner-Heisenberg-Institut\\ 
F\"ohringer Ring 6, 80805 M\"unchen\\ 
Germany}
\end{center}

\begin{center}
{\em $^b$Institut f\"ur Theoretische Physik\\ 
Universit\"at Heidelberg\\ 
Philosophenweg 16, 69120 Heidelberg\\ 
Germany}
\end{center}

\vspace{0.5cm}

\begin{abstract}
\noindent
A detailed error analysis is carried out for the
determination of the \msbar charm quark mass $\overline m_c(\overline
m_c)$ from moments at order $\alpha_s^2$ of the charm cross section in
$e^+e^-$ annihilation. To estimate the theoretical uncertainties the
renormalization scale is implemented in various ways including
energy-dependent functions, which lead to ``contour-improved''
predictions. We obtain $\overline m_c(\overline m_c)=1.29\pm
0.07$~GeV which contains a substantial theoretical uncertainty.

\end{abstract} 

\end{titlepage}

\section*{Introduction}
\label{sectionintroduction}

At the ongoing and future B-physics experiments a realistic estimate
of the uncertainties in the values of the bottom and charm quark
masses will become increasingly important for the measurements of the CKM
parameters and the search for new physics, particularly from inclusive
B-decay rates.~\cite{Battaglia1,Misiak1} Moments of the respective
$e^+e^-$ R-ratios~\cite{Novikov1,Reinders1}, 
\begin{equation}
\label{momdef}
P_n \, = \, \int\frac{ds}{s^{n+1}}\,R_{qq}(s)
\,,
\end{equation}
where $R_{qq}=\sigma(e^+e^-\to q\bar q+X)/\sigma(e^+e^-\to
\mu^+\mu^-)$,
are the most important instrument at present to determine the bottom
and charm quark mass parameters. (See also
Refs.~\cite{Bauer1,Mahmood1} for recent 
bottom quark mass extractions from spectral moments in semileptonic
B-decays.)  

In general, one can distinguish between two regions in $n$, which
require a different theoretical treatment. For low values of $n$ the
moments are dominated by relativistic dynamics and scales of order of the
heavy quark mass $m_q$. This allows the use of the usual expansion in the
number of loops for the theoretical computations, and the \msbar scheme is an
appropriate choice for the heavy quark mass parameter.
However, the lack of data for $R_{qq}$ in the continuum regions
above the quarkonium resonances introduces model-dependent errors that have
not been quantified in most of the previous 
analyses. (See Refs.\,\cite{reviews} for recent reviews.) 
On the other hand, for large values of $n$ the continuum regions are
suppressed and the moments become dominated by the quarkonium
resonance region where good sets of data have been obtained in the
past. However, the theoretical predictions of moments for large values
of $n$ is more complicated  
since the usual loop expansion breaks down and the size of
non-perturbative effects increases. Here, summations of higher
order contributions proportional to powers of $(\alpha_s\sqrt{n})$
need to be carried out in order to capture the relevant 
non-relativistic perturbative information\,\cite{Voloshin1,Hoang1},
and so-called threshold
masses~\cite{Bigithreshmass,Hoangthreshmass,Benekethreshmass} are appropriate
schemes for the heavy quark mass parameter. Moreover, there 
is an upper  ``duality'' bound for the possible choices of $n$ 
since the energy range contributing to the moments, which
is of order $m_q/n$, needs to be larger
than the typical hadronization scale $\Lambda_{\rm
  QCD}$.\,\cite{Hoang1,Poggio1} In the case of $R_{bb}$ and the
determination of the bottom mass, this bound is around $n=10$, and
the two ranges in $n$ are believed to be well separated with their
boundary being approximately at $n=4$. A good number of analyses
exists for small and large values of $n$ and
respecting the cancellation of the dominant renormalon contributions
associated to the choice of the quark mass
definition.\,\cite{Kuhn1,Hoang2,Melnikov1,Hoang3,Beneke1} 
(See also Ref.\,\cite{Jamin1} for an early analysis using the \msbar bottom
quark mass for larger values of $n$.) 

For $R_{cc}$ and the determination of the charm mass, the
distinction between large- and low-$n$ moments is more delicate
because $m_c$ is not much larger than $\Lambda_{\rm QCD}$. Here, the
upper duality bound for $n$ is around $3$ or $4$. This leaves
basically no space at all to carry out the non-relativistic summations
that can be applied in the bottom quark case because the corresponding
techniques are only valid for large $n$ and as long as $m_q/n$ is
larger than $\Lambda_{\rm QCD}$. On 
the other hand, even for $n\leq 4$ the non-relativistic region close
to the $c\bar c$ threshold can have a considerable contribution to the 
moments, while the model-dependences from the experimentally unknown
continuum region can still be significant. For a reliable (error)
analysis these issues need to be taken into account. 

In this paper we determine the \msbar charm quark mass $\overline
m_c(\overline m_c)$ from moments $P_n$ with $n\leq 4$ using the usual
loop expansion in powers of $\alpha_s$. In a previous work~\cite{Kuhn1}
an error of less than 30~MeV was
obtained using results at order $\alpha_s^2$ and a fixed
renormalization scale $\mu$ between $m_c$ and $4m_c$. 
This analysis assumed that the
moments are dominated by scales of order $m_c$. In our analysis we 
focus on the theoretical uncertainties coming from the low-energy
region close to the $c\bar c$ threshold. It is well known that close
to threshold, where the c.m.\ velocity $\beta$ of the quarks is small,
the dominant perturbative contributions to $R_{cc}$ are in fact
governed by  renormalization scales of the order of the relative quark
momentum, $\mu\simeq m_c\beta$.\,\cite{Hoang5,Czarnecki1}
We believe that this effect should not be ignored in determinations of
the charm quark mass since it accounts for important higher-order
information. 
It is the main aim of this work to study the impact of these
higher-order contributions. In particular, we find that they affect 
the estimate of the theoretical error. 
We also carry out a careful study of all sources of
uncertainties including the continuum regions where no data is
available.

\section*{Theoretical Moments}
\label{sectiontheory}

For the QCD parameters used in this work we adopt the \msbar
renormalization scheme and the convention that the charm quark
participates in the running ($n_f=4$). The theoretical moments are directly
parametrized in terms of $\overline m_c\equiv\overline m_c(\overline
m_c)$, and 
the masses of the up, down and strange quarks are set to zero. For the
running of the strong coupling we use four-loop renormalization group
equations, and three-loop matching conditions at the scale $\overline
m_b(\overline m_b)=4.2$~GeV where we switch to $n_f=4$ flavors. We
use three 
different methods to implement the renormalization scale for the
theoretical predictions of the moments and apply order $\alpha_s^2$
results obtained earlier in
Refs.\,\cite{Hoang5,Hoang4,Chetyrkin1,Chetyrkin2}. 

The first method is simply based on a fixed energy-independent
renormalization scale, $\mu^2\simeq m_c^2$, in analogy to previous
analyses. In the OPE including the first non-perturbative contribution
from the dimension-four gluon condensate, the moments have the form     
\begin{equation}
P_n  \, = \, P_n^{\rm pert} + P_n^{\rm non-pert}
\,,
\label{Mn1}
\end{equation}
where
\begin{equation}
P_n^{\rm pert} = 
\frac{1}{( 4\overline m_c^2)^{n}}\,\bigg\{\,
f_n^0 + 
\bigg(\frac{\alpha_s(\mu)}{\pi}\bigg)\,f_n^{1} 
+  
\bigg(\frac{\alpha_s(\mu)}{\pi}\bigg)^2
  \bigg[f_n^{2} +
         \frac{\beta_0}{4}\,f_n^{1}
   \ln\bigg(\frac{\mu^2}{\overline m_c^2}\bigg)
  \bigg]\bigg\}
\end{equation}
and
\begin{eqnarray}
P_n^{\rm non-pert}
& = &
\frac{\langle{\textstyle\frac{\alpha_s}{\pi}}\bmG^2\rangle}
{(4\overline m_c^2)^{2+n}}\,
  \bigg[\,g_n^{0} + 
\bigg(\frac{\alpha_s(\mu)}{\pi}\bigg)\, g_n^{1} 
  \,\bigg]
\,.
\end{eqnarray} 
Here, $\beta_0=11-2/3 n_f$.
The order $\alpha_s$ terms were given in
Refs.\,\cite{Novikov1,Reinders1} and the coefficients for 
the condensate contribution were taken from 
Refs.\,\cite{Broadhurst1,Baikov1} adopting the
renormalization-group-invariant normalization of the
gluon condensate. The
numerical results for the coefficients in Eq.\,(\ref{Mn1}) for
$n=1,2,3,4$ are collected in Tab.\,\ref{tabcoeff1}.
\begin{table}[t!]  
\begin{center}
\begin{tabular}{|c||r|r|r|r|} \hline
$n$ & $1$\mbox{\hspace{6mm}} & $2$\mbox{\hspace{6mm}} & 
      $3$\mbox{\hspace{6mm}} & $4$\mbox{\hspace{6mm}} \\ \hline\hline 
$f_n^{0}$ & $ 1.0667$ & $ 0.4571$ & $ 0.2709$ & $ 0.1847$ \\ \hline 
$f_n^{1}$ & $ 2.5547$ & $ 1.1096$ & $ 0.5194$ & $ 0.2031$ \\ \hline 
$f_n^{2}$ & $ 2.5896$ & $ 2.7790$ & $ 1.6390$ & $ 0.7956$ \\ \hline
$g_n^{0}$ & $-16.042$ & $-26.737$ & $-38.890$ & $-52.352$ \\ \hline 
$g_n^{1}$ & $-15.028$ & $ 13.006$ & $ 78.710$ & $190.935$ \\ \hline 
\end{tabular}
\caption{\label{tabcoeff1} 
Coefficients of the theoretical expressions for the moments $P_n$ at
order $\alpha_s^2$ for massless light quarks based on a fixed
energy-independent renormalization scale. 
}
\end{center}
\vskip 3mm
\end{table}
The numbers for $f_n^0$ and $f_n^1$ agree with results given in
Ref.\,\cite{Kuhn1}. The numbers for $f_n^{2}$ differ because
we also include the contributions from secondary  $c\bar c$
production, where the charm pair is produced through gluon splitting
off primary massless quarks.\,\cite{Hoang4}
For the determination of $\overline m_c(\overline m_c)$ described in
subsequent sections we 
use $\mu^2=\xi^2 M^2$ with $M=1.3$~GeV as the renormalization scale and 
vary the parameter $\xi$ in order to estimate the perturbative
uncertainties. 

\begin{figure}[t] 
\begin{center}
 \leavevmode
 \epsfxsize=6cm
 \leavevmode
 \epsffile[80 475 325 720]{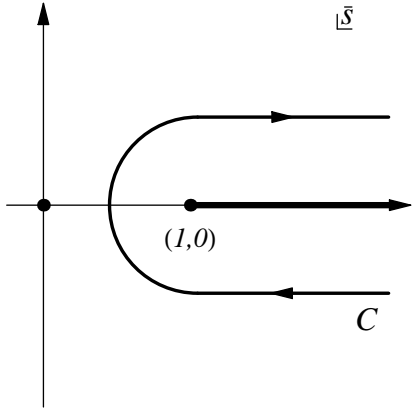}
 \vskip  0.0cm
 \caption{
Path of integration in the complex $\bar s$-plane for the computation
of the moments.
 \label{figcontour} }
\end{center}
\end{figure}

The second method is based on an energy-dependent renormalization
scale of order of the c.m.\ energy, $\mu^2\simeq s$. We note that
the same energy-dependent scale also needs to be used for the
implementation of the \msbar charm mass $\overline m_c(\overline m_c)$ in
order to avoid an incomplete cancellation of the large higher order
corrections associated to the ${\cal O}(\Lambda_{\rm QCD})$ renormalon
of the pole mass definition. A technical issue is related to the fact
that in the \msbar scheme 
for the charm mass parameter, $R_{cc}$ is more singular at the
threshold $s=4\overline m_c$ than in the pole scheme. This makes the
computation of the perturbative contributions of the moments based on
the dispersion integral in Eq.\,(\ref{momdef}) somewhat cumbersome. It is
therefore advantageous to carry out the computation using a contour
integration in the positive complex $s$-plane ($\mbox{Re}(s)>0$) around the
cut of the vacuum polarization function $\Pi_{cc}$ associated to $R_{cc}$,
\begin{eqnarray}
\label{momdefcomplex}
P_n & = & 
\frac{1}{2\pi i}
\int_{\cal C}\frac{ds}{s^{n+1}}\,\Pi_{cc}\bigg(\frac{s}{4\overline m_c^2}\bigg)
\nonumber\\ 
& = &
\frac{1}{(4\overline m_c^2)^n}\,
\frac{1}{2\pi i}
\int_{\cal C}\frac{d\bar s}{\bar s^{n+1}}\,\Pi_{cc}(\bar s)
\,.
\end{eqnarray}
The same procedure can of course also be applied for the first method.
The path ${\cal C}$ of the integration is illustrated in
Fig.\,\ref{figcontour}. 
%
%
For practical purposes we use $\bar s\equiv s/(4\overline m_c)$ as the
integration variable and $\mu^2=\xi^2 M^2 \bar s$ as the
renormalization scale. This choice allows to
explicitly factor out the dependence of the moments on $\overline m_c$. The
perturbative part of the moments can then be written in the form
\begin{eqnarray}
P_n^{\rm pert} & = &
\frac{1}{(4\overline m_c^2)^{n}}\,\bigg\{\,
f_n^0 + C_n^1 +
\bigg[ C_n^{20} + \frac{\beta_0}{4} \bigg(
C_n^{21} \ln\bigg(\frac{\xi^2 M^2}{\overline m_c^2}\bigg) + C_n^{22}
\bigg)\bigg]\bigg\}
\,,
\end{eqnarray}
where the superscripts $0,1,2$ refer to ${\cal O}(1)$, 
${\cal O}(\alpha_s)$ and ${\cal O}(\alpha_s^2)$,
respectively. Exemplarily, for $\xi=1$ ($\xi=2$) and 
$\alpha_s^{n_f=5}(M_z)=0.118$
(i.e.\ $\alpha_s^{n_f=4}(4.2\,\mbox{GeV})=0.225$), the coefficients are given
numerically in Tab.\,\ref{tabcoeff23}
\begin{table}[t!]  
\begin{center}
\begin{tabular}{|c||c|c|c|c|} \hline
$n$ & 1 & 2 & 3 & 4 \\ \hline\hline 
$C_n^{1}$ & $ 0.2433$ ($0.1839$) & $ 0.1093$ ($0.0822$) 
          & $ 0.0498$ ($0.0379$) & $ 0.0167$ ($0.0135$)\\ \hline 
$C_n^{20}$ & $ 0.0434$ ($0.0223$) & $ 0.0308$ ($0.0169$) 
           & $ 0.0165$ ($0.0094$) & $ 0.0071$ ($0.0042$)\\ \hline 
$C_n^{21}$ & $ 0.0235$ ($0.0134$) & $ 0.0107$ ($0.0061$)
           & $ 0.0047$ ($0.0027$) & $ 0.0012$ ($0.0009$)\\ \hline 
$C_n^{22}$ & $ 0.0198$ ($0.0116$) & $ 0.0082$ ($0.0046$)
           & $ 0.0044$ ($0.0024$) & $ 0.0027$ ($0.0015$)\\ \hline\hline 
$\tilde C_n^{1}$ & $ 0.3105$ ($0.2183$) & $ 0.2339$ ($0.1400$) 
                 & $ 0.1917$ ($0.0968$) & $ 0.1654$ ($0.0685$)\\ \hline  
$\tilde C_n^{20}$ & $ 0.0385$ ($0.0190$) & $ 0.0658$ ($0.0283$) 
                  & $ 0.0759$ ($0.0270$) & $ 0.0820$ ($0.0239$)\\ \hline 
$\tilde C_n^{21}$ & $ 0.0377$ ($0.0187$) & $ 0.0405$ ($0.0158$)
                  & $ 0.0428$ ($0.0136$) & $ 0.0455$ ($0.0117$)\\ \hline 
$\tilde C_n^{22}$ & $ 0.0000$ ($0.0000$) & $-0.0377$ ($-0.0187$) 
                  & $-0.0593$ ($-0.0252$)& $-0.0756$ ($-0.0277$)\\ \hline
\end{tabular}
\caption{\label{tabcoeff23} 
Coefficients of the theoretical expressions for the moments $P_n$ at
order $\alpha_s^2$ for massless light quarks based on 
energy-dependent renormalization scales as described in the text for
$\xi=1$ ($\xi=2$).
}
\end{center}
\vskip 3mm
\end{table}

The third method is based on an energy-dependent renormalization scale of
order of the relative three-momentum of the charm quarks, $\mu^2\simeq
-s\beta^2$, where $\beta\equiv\sqrt{1-4\overline m_c^2/s}$. Again, the
perturbative 
part of the moments is computed using the contour integration in
Eq.\,(\ref{momdefcomplex}). In the complex $\mu^2$-plane, $\alpha_s(\mu^2)$
has a cut along the negative real axis. It is therefore necessary to introduce
the negative sign. With this choice the cut in
$\alpha_s(\mu^2\simeq -s\beta^2)$ agrees with the one of the vacuum
polarization function, and contours can be found such that $\beta$ is never
small along the path, and the
perturbative description is appropriate.\footnote{
For the first and the third method one can
also choose a closed path around the origin in the counter-clockwise
direction.  
} 
In fact, this is in close 
analogy to the so-called ``contour-improved'' approach to compute the hadronic
$\tau$-decay width (see e.g.\ Ref.\,\cite{Pich1,Pivovarov1}). For our
numerical computations we use $\bar s=s/(4\overline m_c^2)$ as the
integration variable and $\mu^2=\xi^2 M^2(1-\bar s)$ as the
renormalization scale. The perturbative part of the moments can then
be written in the form 
\begin{eqnarray}
P_n^{\rm pert} & = &
\frac{1}{(4\overline m_c^2)^{n}}\,\bigg\{\,
f_n^0 + \tilde C_n^1 +
\bigg[ \tilde C_n^{20} + \frac{\beta_0}{4} \bigg(
\tilde C_n^{21} \ln\bigg(\frac{\xi^2 M^2}{\overline m_c^2}\bigg) + \tilde C_n^{22}
\bigg)\bigg]\bigg\}
\,.
\end{eqnarray}
Exemplarily, for $\xi=1$ ($\xi=2$) and $\alpha_s^{n_f=5}(M_z)=0.118$
(i.e.\ $\alpha_s^{n_f=4}(4.2\,\mbox{GeV})=0.225$), the coefficients are given
in Tab.\,\ref{tabcoeff23}. Note that $\tilde C_1^{22}=0$ for any choice
of the parameters because the corresponding integrand does not have a pole at
$\bar s=0$ due to the additional factor of $\ln(1-\bar s)$.

\section*{Experimental Moments}
\label{sectionexperiment}

For the contributions to the experimental moments from the
$J/\psi$ and the $\psi^\prime$ we use the most recent averages for
masses and $e^+e^-$ widths given by the PDG~\cite{PDG}
($M_{J/\psi}=3.0969$~GeV, 
$M_{\psi^\prime}=3.6860$~GeV,
$\Gamma^{e^+e^-}_{J/\psi}=5.14\pm 0.31$~keV,
$\Gamma^{e^+e^-}_{\psi^\prime}=2.12\pm 0.12$~keV). 

For the energy region between $3.73$ and $4.8$~GeV we employ the data
for the total R ratio, $R_{\rm tot}$, obtained by the BES
collaboration.\,\cite{BES} To obtain experimental data for $R_{cc}$ we
use the last four data points below $3.73$~GeV 
to fit the non-charm R ratio, $R_{\rm nc}$, assuming it
to be energy-independent, and subtract it from the numbers for 
$R_{\rm tot}$. The effects from the Z exchange and the
small logarithmic energy-dependence expected from theory are below 
1\% and can be neglected.
The determination of the statistical uncertainties of $R_{cc}$ is
standard. The determination of the systematical uncertainties in
$R_{cc}$ is more subtle because only systematical errors for $R_{\rm
  tot}$ (with all individual contributions being added quadratically)
but no specific numbers for the D-meson final states above $3.7$~GeV
were given in Ref.\,\cite{BES}. Thus, naively subtracting the
systematical errors of $R_{\rm nc}$ from the ones in $R_{\rm tot}$
underestimates the actual systematical errors in $R_{\rm cc}$. To
obtain numbers for the systematical errors in $R_{\rm cc}$ we make the
assumption that the systematical errors for the non-charm and the
charm final states are equally large, i.e. the systematical errors
from $R_{\rm cc}$ are obtained by multiplying those of $R_{\rm tot}$
by a factor $1/\sqrt{2}$. This procedure is supported by the
observation that the relative systematical errors given in
Ref.\,\cite{BES} for energies above $3.7$~GeV are in average
approximately a factor $1.5$ larger than expected from extrapolating
the relative systematical errors for energies below $3.7$~GeV.

In the region above $4.8$~GeV the only other data available for
$R_{\rm tot}$ is from the MD1 experiment~\cite{MD1} and from the CLEO
collaboration~\cite{CLEO1} in the energy region between $7.25$ and
$10.5$~GeV. For the determination of $R_{\rm cc}$ and the
uncertainties we use the same method as described in the previous
paragraph. The resulting relative systematical errors amount to about
8\%. 

There is no continuous experimental data for $R_{\rm cc}$ in the
regions between
$4.8$ and $7.25$~GeV and above $10.5$~GeV. As shown below, these
contribution turn out to be not very important. 
For our analysis we use a model for these contributions based on the
theory prediction of $R_{\rm cc}$ at order $\alpha_s^2$. This is
reasonable since it is known that perturbative QCD agrees with data
for $c\bar c$ production to several percent at $M_Z$ and at the level
of a few $10\%$ percent in the LEP2 region.\,\cite{LEP2}. The resulting
model-dependence in the determination of $\overline m_c$ is negligible
for $n>1$, as shown below. For $n=1$ the 
contribution from the region between $4.8$ and $7.25$~GeV leads to a
model-dependent error of about $10$~MeV. 

\begin{table}[t!]  
\begin{small}
\begin{center}
\begin{tabular}{|c||c|c|c|c|} \hline
 & $P_1$ & $P_2$ & $P_3$ & $P_4$  \\
\raisebox{1.5ex}[-1.5ex]{contribution} &
 $\times\,10^2~\mbox{GeV}^2$ & $\times\,10^3~\mbox{GeV}^4$ &
 $\times\,10^4~\mbox{GeV}^6$ & $\times\,10^5~\mbox{GeV}^8$ \\
\hline\hline
$J/\psi$&$  8.82(38|38)$&$  9.20(39|39)$&$  9.59(41|41)$&$ 10.00(43|43)$
\\\hline
$\psi^\prime$&$  2.16(9|9)$&$  1.59(6|6)$&$  1.17(5|5)$&$  0.86(3|3)$
\\\hline
$3.7-4.8~\mbox{GeV}$&$  3.23(6|33)$&$  1.80(3|19)$&$  1.02(2|11)$&$  0.59(1|7)$
\\\hline
$4.8-7.25~\mbox{GeV}$&$  3.82(25)$&$  1.19(8)$&$  0.39(3)$&$  0.13(1)$
\\\hline
$7.25-10.5~\mbox{GeV}$&$  1.45(3|10)$&$  0.21(0|1)$&$  0.03(0|0)$&$  0.00(0|0)$
\\\hline
$10.5~\mbox{GeV}-M_Z$&$  1.27(1)$&$  0.06(0)$&$  0.00(0)$&$  0.00(0)$
\\\hline
$M_Z-\infty$&$  0.02(0)$&$  0.00(0)$&$  0.00(0)$&$  0.00(0)$
\\\hline
\end{tabular}
\caption{\label{tabdatamom}
Individual contributions to the experimental moments including statistical and
systematical uncertainties. If only one error is given, it comes from theory.
The resonance contributions have been determined in the narrow width
approximation,
$(P_n)_k=9\pi\,\Gamma^{e^+e^-}_k/[\alpha^2(3.1\,\mbox{GeV})\,M_k^{2n+1}]$,
where for the electromagnetic coupling
$[\alpha(3.1~\mbox{GeV})]^{-1}=134.3$ has been adopted.
}
\end{center}
\end{small}
\vskip 3mm
\end{table}

In Tab.\,\ref{tabdatamom} a collection of all contributions to the
first four moments is given. For contributions coming from 
experimental data the first error is statistical and the second
systematical. For the contributions from the $J/\psi$ and the
$\psi^\prime$ we used the averaged PDG~\cite{PDG} errors. Since detailed
numbers for the various error sources are difficult to
obtain, we make the simplified assumption that statistical errors
(being later combined in quadrature) and systematical errors (being
later combined linearly) are equal. This is consistent with
information given in the original literature. For the 
continuum regions that have to be estimated from our theory-model the
contributions from below and above $M_Z$ are distinguished in order to
visualize the impact of these two regions. The respective errors given
in Tab.\,\ref{tabdatamom} come from varying the strong coupling in the
range $\alpha_s(M_Z)=0.118\pm 0.003$, the \msbar charm mass in the
conservative range $\overline m_c=1.3\pm 0.2$~GeV and the
renormalization scale between $\mu^2=s/16$ and $\mu^2=s$. The central
values are obtained from the average of the corresponding extremal
values. The relative error obtained in this way is not larger than 7\%
for all three regions where theory is employed. Note that for our error
estimates for $\overline m_c$ (see Tab.\,\ref{tabmcerrors}) we adopt, by hand, 
a relative error of 10\%. This reduces the model-dependence of the $n=1$
results to an acceptable level.

\section*{Uncertainties in $\overline m_c(\overline m_c)$}
\label{sectionmm}

\begin{table}[t!]  
\begin{small}
\begin{center}
\begin{tabular}{|c||c|c|c|c|} \hline
$n$ & 1 & 2 & 3 & 4  \\
\hline\hline
Method 1 $(3/4\leq\xi\leq 3)$ 
& $1.286(100)$ & $1.280(66)$ & $1.274(51)$ & $1.269(44)$
\\ \hline
Method 2 $(3/4\leq\xi\leq 3)$ 
& $1.280(101)$ & $1.272(61)$ & $1.267(47)$ & $1.264(41)$
\\ \hline
Method 3 $(3/4\leq\xi\leq 3)$ 
& $1.287(100)$ & $1.296(69)$ & $1.298(58)$ & $1.298(53)$
\\ \hline
combined $(3/4\leq\xi\leq 3)$ 
& $1.283(104)$ & $1.288(77)$ & $1.288(68)$ & $1.287(64)$\\
\hline\hline
Method 1 $(1\leq\xi\leq 4)$ 
& $1.277(94)$ & $1.272(59)$ & $1.267(47)$ & $1.264(45)$
\\ \hline
Method 2 $(1\leq\xi\leq 4)$ 
& $1.268(92)$ & $1.264(56)$ & $1.262(45)$ & $1.260(43)$
\\ \hline
Method 3 $(1\leq\xi\leq 4)$ 
& $1.278(94)$ & $1.291(65)$ & $1.294(55)$ & $1.295(52)$
\\ \hline
combined $(1\leq\xi\leq 4)$ 
& $1.274(98)$ & $1.282(74)$ & $1.283(66)$ & $1.282(65)$
\\ \hline
\end{tabular}
\caption{\label{tabmcfinal}
Results for $\overline m_c(\overline m_c)$ for different methods to
implement the renormalization scale. The uncertainties represent the sum of
all individual error sources. The combined result represents the total mass
range covered by the three methods.
All numbers are given in units of GeV.
}
\end{center}
\end{small}
\vskip 3mm
\end{table}

For the determination of $\overline m_c(\overline m_c)$ we fit single
moments using the three different methods for the theoretical predictions
described above. The contribution from the gluon
condensate is taken from the first method for all cases.
We use $\alpha_s^{n_f=5}(M_z)=0.118\pm 0.003$
($\alpha_s^{n_f=4}(4.2\,\mbox{GeV})=0.225\pm 0.012$) and  
$\langle{\textstyle\frac{\alpha_s}{\pi}}\bmG^2\rangle
=(0.024\pm 0.024)~\mbox{GeV}^4$ as theoretical input.
Our final results for $n=1,2,3,4$, for the three different theoretical
methods and for varying the renormalization scale parameter in the
ranges $3/4\leq\xi\leq 4$ and $1\leq\xi\leq 4$ are given in
Tab.\,\ref{tabmcfinal}. The central values are obtained using the
respective central values for all experimental data and the
theoretical input parameters, and by taking the median of the range in
$\overline m_c(\overline m_c)$ covered by varying $\xi$ in the
respective ranges. The uncertainties are obtained by combining all individual
error sources as described below and shown in Tab.\,\ref{tabmcerrors}.
In Tab.\,\ref{tabmcfinal} we have also given combined results 
representing the total range in $\overline m_c$
covered by the three theoretical methods.

\begin{table}[t!]  
\begin{center}
\begin{tabular}{|c||c|c|c|c||c|c|c|c|} \hline
& \multicolumn{4}{|c||}{Method 2} & 
   \multicolumn{4}{|c|}{Method 3}
\\ \hline
$n$ & 1 & 2 & 3 & 4 & 1 & 2 & 3 & 4 
\\ \hline\hline
\multicolumn{9}{|c|}{$3/4\leq\xi\leq 3$}
\\ \hline
central & 1280 & 1272 & 1267 & 1264 & 1287 & 1296 & 1298 & 1298
\\ \hline
$J/\psi$ & 12/12 & 9/9 & 7/7 & 6/6 & 11/11 & 9/9 & 7/7 & 6/6
\\ \hline
$\psi^\prime$ & 3/3 & 1/1 & 1/1 & 0/0 & 3/3 & 1/1 & 1/1 & 0/0
\\ \hline
$3.73 - 4.8$~GeV & 2/10 & 1/4 & 0/2 & 0/1 & 2/10 & 1/4 & 0/2 & 0/1
\\ \hline
$4.8 - 7.25$~GeV & 12 & 3 & 1 & 0 & 11 & 3 & 1 & 0
\\ \hline
$7.25 - 10.5$~GeV & 1/3 & 0/0 & 0/0 & 0/0 & 1/3 & 0/0 & 0/0 & 0/0 
\\ \hline
$10.5$~GeV $-\,\infty$ & 4 & 0 & 0 & 0 & 4 & 0 & 0 & 0
\\ \hline
$\delta \langle{\textstyle\frac{\alpha_s}{\pi}}\bmG^2\rangle$
& 4 & 8 & 11 & 14 & 4 & 6 & 8 & 9
\\ \hline
$\delta\alpha_s(M_Z)$ 
& 14 & 8 & 5 & 3 & 17 & 17 & 17 & 18
\\ \hline
$\delta\xi$ 
& 28 & 18 & 12 & 9 & 25 & 19 & 15 & 12
\\ \hline
combined error 
& 101 & 61 & 47 & 41 & 100 & 69 & 58 & 53
\\ \hline\hline
\multicolumn{9}{|c|}{$1\leq\xi\leq 4$}
\\ \hline
central & 1268 & 1264 & 1262 & 1260 & 1278 & 1291 & 1294 & 1295
\\ \hline
$J/\psi$ & 12/12 & 9/9 & 7/7 & 6/6 & 12/12 & 9/9 & 7/7 & 6/6
\\ \hline
$\psi^\prime$ & 3/3 & 1/1 & 1/1 & 0/0 & 3/3 & 1/1 & 1/1 & 0/0
\\ \hline
$3.73 - 4.8$~GeV & 2/10 & 1/4 & 0/2 & 0/1 & 2/10 & 1/4 & 0/2 & 0/1
\\ \hline
$4.8 - 7.25$~GeV & 12 & 3 & 1 & 0 & 12 & 3 & 1 & 0
\\ \hline
$7.25 - 10.5$~GeV & 1/3 & 0/0 & 0/0 & 0/0 & 1/3 & 0/0 & 0/0 & 0/0 
\\ \hline
$10.5$~GeV $-\,\infty$ & 4 & 0 & 0 & 0 & 4 & 0 & 0 & 0
\\ \hline
$\delta \langle{\textstyle\frac{\alpha_s}{\pi}}\bmG^2\rangle$
& 5 & 9 & 14 & 20 & 5 & 7 & 10 & 14
\\ \hline
$\delta\alpha_s(M_Z)$ 
& 11 & 5 & 3 & 2 & 11 & 10 & 9 & 9
\\ \hline
$\delta\xi$ 
& 24 & 15 & 10 & 7 & 24 & 20 & 18 & 14
\\ \hline
combined error 
& 92 & 56 & 45 & 43 & 94 & 65 & 55 & 52
\\ \hline\hline
\end{tabular}
\caption{\label{tabmcerrors} 
Central values and individual errors for $\overline m_c(\overline m_c)$ in
units of MeV based on the methods described in the text. 
}
\end{center}
\vskip 3mm
\end{table}

A detailed account of individual errors is presented in
Tab.\,\ref{tabmcerrors}. For experimental data the first error is
statistical and the second systematical. The results are only given
for the second and the third theoretical method, as the first and
second methods yield very similar results, particularly for $n>1$. This is
expected since the high-energy range, $s\gg 4\overline m_c^2$, where the 
renormalization scales of the two methods differ, is suppressed. The
uncertainties from experimental data and from the treatment of
the non-perturbative corrections are
similar for the three theoretical methods. As mentioned before,
the model-dependent error coming from the lack of experimental data in the 
continuum region is negligible for $n>1$ and amounts to only about
10~MeV for $n=1$. This is in  contrast to determinations of $\overline
m_b(\overline m_b)$ where the model-dependent error is significantly
larger, particularly for $n=1$ and $n=2$.~\cite{Hoang2} 
The uncertainties coming from the error in $\alpha_s(M_Z)$ for $n>1$ are larger for
the third method than for the first and the second one, since variations in
the value of $\alpha_s(M_Z)$ have a larger impact at lower scales. 
To obtain combined errors we have
added all statistical uncertainties in quadrature, and all other
uncertainties linearly. Since the statistical errors are smaller than
the other sources of uncertainties, the reader should note that our total
errors cannot be interpreted statistically.

It is instructive to have a closer look onto the perturbative
uncertainties obtained by the different theoretical methods. For all
methods the uncertainties obtained by varying $\xi$
either in the range $3/4\leq\xi\leq 3$ or $1\leq\xi\leq 4$ become
smaller for larger values of $n$. This behavior is, however, less pronounced
for the third method. 
Thus, each of the methods yields a behavior in contrast to the expectation 
that the scaling uncertainties grow like the uncertainties from the
non-perturbative corrections, since for larger $n$ the moments become
dominated by dynamics at lower energies. 
On the other hand, comparing the results for the central
values obtained by the three theoretical methods it is found that
they tend to decrease for larger values of $n$ for the first and
second method, while there is the opposite effect for the third
method. In fact, the difference between the central values for $n>1$ 
is larger than the uncertainties based on varying $\xi$.
This feature is related to the different treatment of
higher-order contributions and appears to be similar to the 
discrepancies found between the ``contour-improved'' and the
``fixed-scale'' approaches for predictions of the hadronic
$\tau$-decay rate. (See e.g. Refs.\,\cite{Groote1,Baikov2} for discussions on 
this issue). Interestingly, the convergence behavior 
of the moments for all three methods is quite good. For example, for
$n=3$, $\overline m_c=1.3$~GeV, $\xi=2$ and $\alpha_s(M_Z)=0.118$ one
obtains (in units of $10^{-4}$~GeV$^{-6}$)   
$8.77+1.44+0.75=10.95$,
$8.77+1.23+0.72=10.72$ and
$8.77+3.13+0.44=12.35$ 
for the perturbative expansion of the moment
for the first, second and the third method, respectively. The
discrepancy between the first two methods and the third one is
considerably larger than an error estimate based on the 
size of the order $\alpha_s^2$ terms obtained from each of the methods.
These discrepancies represent theoretical
uncertainties that are not captured by conventional variations in 
the renormalization scale. However, they should be included in the estimate of
the theoretical error. We include this 
type of theoretical uncertainty in our procedure to determine the  
combined results shown in Tab.\,\ref{tabmcfinal}.

As our final result we adopt
\begin{equation}
\overline m_c(\overline m_c) \, = \, 1.29\pm 0.07~\mbox{GeV}
\end{equation} 
obtained from our fits using the moments for $n=2,3$. They are sufficiently below
the duality bound and, at the same time, have only small sensitivity to
the continuum region above the $\psi^\prime$ resonance.
Our result is compatible with earlier analyses based on moments of
$R_{cc}$~\cite{Kuhn1,Eidemuller1,Erler1,Ioffe1,Eidemuller2}, and with 
recent lattice determinations~\cite{Becirevic1,Juge1,Rolf1} as well as with
the finite energy sum rule analysis of Ref.\,\cite{Penarrocha1}. 
However, our error is much larger than in 
Refs.\,\cite{Kuhn1,Erler1,Ioffe1} due to our more pertinent treatment of
theoretical and experimental uncertainties. 
On the other hand, the size of our error is comparable 
to Refs.\,\cite{Eidemuller1,Eidemuller2} where summations of terms 
proportional to powers of $(\alpha_s\sqrt{n})$~\cite{Voloshin1,Hoang1}
were partially included to account for non-relativistic effects in the
low-energy region $s\simeq 4m_c^2$. However, the resulting central values are 
systematically lower than ours by up to $100$~MeV. This effect
apparently arises from the summation of the higher order terms just
mentioned and from the use of moments for $n>4$. Both issues are
problematic from the conceptual point of view. 
The size of our error is also comparable to the analysis of
Ref.\,\cite{Penarrocha1}, but our central value is about $100$~MeV lower.
This discrepancy seems to indicate that there is a systematic difference
between the sum rules based on Eq.\,(\ref{momdefcomplex}) and the 
finite energy moments employed in Ref.\,\cite{Penarrocha1}. We note, however,
that the strong influence of systematic errors in the derivation of the
non-charm R ratio $R_{\rm nc}$ and possible ambiguities in the choice of the
upper energy cut-off represent sources of uncertainties for the finite energy
sum rule approach that are not easy to quantify. 

Based on the numbers given in Tabs.\,\ref{tabmcfinal} and
\ref{tabmcerrors} it is straightforward to discuss how the uncertainty
in $\overline m_c(\overline m_c)$ might be further reduced in the
future. One improvement could be achieved by the
computation of the order $\alpha_s^3$ corrections to the moments -- if
they indeed lead to a reduction of the discrepancies between
``contour-improved'' and ``fixed-scale'' predictions. At present the
theoretical uncertainty 
coming from the different ways to implement the renormalization scale
amounts to about $30$ to $35$~MeV. The uncertainties coming from the $J/\psi$  
$e^+e^-$-partial width, from the error in $\alpha_s$ and from the
treatment of non-perturbative effects are each at the level of
$10$~MeV. For the combination of these uncertainties a reduction of up
to $15$~MeV appears possible for an improved measurement of
the $e^+e^-$-partial width and if smaller uncertainties in the
determinations of the strong coupling are achieved. On the other
hand, improvements in the data for $\psi^\prime$ and for the $c\bar c$
continuum will not improve the situation significantly. 

We have also applied our method to estimate the theoretical
uncertainties for determinations of $\overline m_b(\overline m_b)$
from low-$n$ moments of $R_{bb}$. Using the approach described above
we find $\overline m_b(\overline m_b)=4.22\pm 0.11$~GeV. The error is
only $20$~MeV larger than the error obtained from an earlier analysis
by one of the authors which was based only the first method with an
energy-independent renormalization scale. This
shows that the scale choice $\mu^2\simeq -s\beta^2$ has a smaller
impact in the bottom quark case due to its larger mass.

\vspace{1cm}

\bibliographystyle{prsty}

\end{document}